\documentclass[12pt]{article}
\usepackage{color}
\usepackage{amsmath,amssymb}
\usepackage[dvipdfm]{graphicx}
\usepackage[dvipdfm]{graphicx}
\usepackage{cancel}

\begin{document}
\baselineskip=16pt
\begin{titlepage}
\begin{flushright}
{\small OU-HET 696/2011}\\
{\small KU-PH-008}\\
\end{flushright}
\vspace*{1.2cm}
\begin{center}

{\Large\bf 
Tiny neutrino mass from SUSY and lepton number breaking sector 
} 
\lineskip .75em
\vskip 1.5cm

\normalsize
$^1${\large Naoyuki Haba}, 
and
$^2${\large Tetsuo Shindou}

\vspace{1cm}

$^1${\it Department of Physics, 
 Osaka University, Toyonaka, Osaka 560-0043, 
 Japan} \\

$^2${\it Faculty of Engineering, Kogakuin University, Tokyo 163-8677, Japan}

\vspace*{10mm}

{\bf Abstract}\\[5mm]
{\parbox{13cm}{\hspace{5mm}
%

We suggest new setup where 
 SUSY breaking spurion $F$-term possesses lepton number.
This setup not only modifies sparticle mass spectra 
 but also realizes several new models, 
 where neutrino mass 
 is naturally induced through radiative corrections. 
We here suggest two new models; the first one is 
 (i): pseudo-Dirac/Schizophrenic neutrino model, and 
 the second one is  
 (ii): pure Majorana neutrino model.
We will also show 
 this setup can naturally apply to the supersymmetric 
 Zee-Babu model.

}}

\end{center}
\end{titlepage}

\section{Introduction}

Recent neutrino oscillation experiments 
 gradually reveal a 
 structure of 
 lepton sector\cite{Strumia:2006db, analyses}.  
However, an origin of generation structure 
 is still a mystery, and we do not know a mechanism of 
 constructing 
 generation structure in lepton sector as well as 
 quark sector. 
Thus, 
 finding a mechanism of  
 small neutrino mass and lepton 
 flavour structure
 is one of the greatest 
 keys of revealing new physics beyond the Standard Model (SM). 
One interesting idea is that tiny neutrino mass is realized through
radiative corrections\cite{zee}. Where quantum corrections induce effective
dimension five operator, and Majorana neutrino mass is obtained. The
smallness of neutrino mass is natural because it is induced by quantum
corrections. There are some models where tiny Majorana neutrino mass
is induced through one-loop diagram\cite{zee,one-loop}, two-loop\cite{zee-babu,two-loop,Aoki:2010ib}, 
three-loop\cite{three-loop}, and so on.
Anyhow, they have new extra particles, and 
 induce new phenomenology beyond the SM, hence 
 experimental evidence is expected in LHC.

On the other hand, supersymmetry (SUSY) is 
 expected to be the most promising candidate of 
 physics beyond the SM. 
There are rich phenomenology such as 
 gauge coupling unification, existence of 
 dark matter, and so on. 
However, experiments have not proved its existence yet, 
 and the SUSY 
 breaking mechanism is still a mystery. 
Thus, it is an important research to 
 reveal an origin of SUSY breaking. 
Spurion field is sometimes introduced 
 as an auxiliary field which has only $F$-term 
 to represent SUSY breaking mass parameters.   
The origin of spurion field is considered as 
 super-heavy particle in a underlying theory, 
 and then it has no vacuum expectation 
 value (VEV) of scalar component\footnote{
The authors thank S.~Yamaguchi for private discussions. 
}.

In this paper, 
 we suggest new setup where 
 SUSY breaking spurion $F$-term possesses lepton number.
This setup not only modifies sparticle mass spectra 
 but also realizes several new models, 
 where neutrino mass 
 is naturally 
 induced by radiative corrections. 
We here suggest two new models; the first one is 
 (i): pseudo-Dirac/Schizophrenic neutrino model, and 
 the second one is  
 (ii): pure Majorana neutrino model.
We will also show 
 this setup can naturally apply to the supersymmetric 
 Zee-Babu model.
The necessary lepton number violating couplings 
 are induced in the soft SUSY breaking interactions, 
 while the lepton number is kept in the superpotential.

\section{Spurion fields and SUSY breaking terms}

Let us consider a setup where 
SUSY breaking $F$-term possesses lepton number. 
We introduce a hidden sector, where SUSY is broken 
by two types of gauge singlet spurion fields.
One type of the spurions has no lepton number,
and the other type has a lepton number.
In the following we denote the lepton number singlet spurion as $X$, 
and the spurion with a lepton number as $X_L^{}$.
We assume these spurion fields have non-zero $F$-terms,
 whose effects are 
 transformed to our visible sector as soft SUSY breaking terms 
 through gravity interactions. 
The spurion fields are expected to have no VEV at scalar 
 component. 
Of cause, we can also consider other messenger of SUSY breaking,
 such as 
 gauge mediation, 
 but here we would like to show the gravity mediation, for simplicity. 
We consider gauge singlet spurion fields and impose 
 $R$-parity conservation, hereafter. 

\begin{table}[h]
\caption{Matter contents of the MSSM. 
\label{table:model}}
\begin{center}
\begin{tabular}{|c||c|c|c|c|}\hline
Superfields&$\text{SU(3)}_{\text{C}}^{}$&$\text{SU(2)}_{\text{L}}^{}$&$\text{U(1)}_{Y}^{}$&$\text{U(1)}_{L}^{}$\\\hline
$Q_i$&$\pmb{3}$&$\pmb{2}$&$+\frac{1}{6}$&$0$\\
$U_i^c$&$\pmb{3}$&$\pmb{1}$&$-\frac{2}{3}$&$0$\\
$D_i^c$&$\pmb{3}$&$\pmb{1}$&$+\frac{1}{3}$&$0$\\
$L_i$&$\pmb{1}$&$\pmb{2}$&$-\frac{1}{2}$&$1$\\
$E_i^c$&$\pmb{1}$&$\pmb{1}$&$+1$&$-1$\\
$H_1^{}$&$\pmb{1}$&$\pmb{2}$&$-\frac{1}{2}$&$0$\\
$H_2^{}$&$\pmb{1}$&$\pmb{2}$&$+\frac{1}{2}$&$0$\\ \hline
\end{tabular}
\end{center}
\end{table}

Usually, only $X$ is considered in order to give soft SUSY breaking terms
in the MSSM, whose fields content is shown in Table~\ref{table:model}.
The superpotential of the MSSM is given by
\begin{align}
W^{\text{MSSM}}=&
 y^u_{ij} U_i^cQ_j\cdot H_2^{}+
 y^d_{ij} D_i^cQ_j\cdot H_1^{}+
 y^e_{ij} E_i^cL_j\cdot H_1^{}+
 \mu H_1\cdot H_2 . 
\end{align}
In the MSSM with spurion $X$,
 the soft SUSY breaking terms are induced as
\begin{align}
\mathcal{L}_{\text{MSSM}}^{\rm soft}=&
\int d^4\theta\left\{
	\frac{(c_\Phi^{})^{ij}}{M^2}X^{\dagger}X
	\Phi^{\dagger}_i \Phi_j^{}
	\right\}
	\nonumber\\
	&+\int d^2\theta\left\{
		\frac{(a_{U}^{})^{ij}}{M}XU_i^cQ_i\cdot H_2^{}
		+\frac{(a_{D}^{})^{ij}}{M}XD_i^cQ_j\cdot H_1^{}\nonumber\right.\\
	&\left.\phantom{+\int d^2\theta()()()}
		+\frac{(a_{E}^{})^{ij}}{M}XE_i^cL_j\cdot H_1^{}
		+b_H^{}XH_1\cdot H_2^{}
		+\text{h.c.}
		\right\}\nonumber\\
	&+\int d^2\theta\left\{
		\frac{(d_{3}^{})^{}}{M}XG^\alpha G_\alpha 
		+\frac{(d_{2}^{})^{}}{M}XW^\alpha W_\alpha 
		+\frac{(d_{1}^{})^{}}{M}XB^\alpha B_\alpha 
		+\text{h.c.}
		\right\}\;,
\end{align}
where we denote the SM matter and Higgs fields as
 $\Phi_i (=Q_i, L_i, D_i^c, U_i^c, E_i^c, H_1, H_2)$. 
When the $F$-term of $X$, which is denoted as $F_X$, 
 has non-zero magnitude,
 the first line of the above equation induces the scalar masses, 
 the second and third lines induce
 the tri-linear scalar couplings, and the last line gives the gaugino masses.

In addition to ordinal soft SUSY breaking terms induced by 
 $F_X$, we can have SUSY breaking effects from 
 $F$ components of $X_L^{}$, denoted as $F_{X_L}$. 
In the hidden sector, not only the SUSY but also the lepton number is
 softly broken by the $F_{X_L}$. 
Additional terms to $\mathcal{L}_{\text{MSSM}}^{\rm soft}$ induced from the
interactions between spurion field $X_L$ 
 and the MSSM fields can be written as
\begin{align}
\mathcal{L}_{\text{MSSM-}\not{L}}^{\rm soft}=&
\mathcal{L}_{\text{MSSM}}^{\rm soft}+
\int d^4\theta\left\{
	\frac{(\tilde{c}_\Phi^{})^{ij}}{M^2}X_L^{\dagger}X_L^{}
	\Phi^{\dagger}_i \Phi_j^{}\right\}\;.
\end{align}
Notice that $F_X$ induces all soft masses, while 
 $F_{X_L}$ does not induce $A$-term, $B$-term,
 and gaugino mass terms. 
Thus,  $F_{X_L}$ only contributes to soft ``squared''
 masses due to the lepton number symmetry. 
In other words, 
 $F_{X_L}$ can induce non-universal effects 
 with $F_{X_L} < F_X$. 
Anyhow, we show that 
 this setup can suggest new models.

\section{New models}

In the framework with $X_L^{}$,
 new radiative induced neutrino mass models can be
 considered.
Let us show three examples of such models.
In these models, the lepton number is conserved in the superpotential, 
 but it is {\it softly} 
 broken in the SUSY breaking interactions. 

\subsection{Loop induced Majorana neutrino mass model I}

The first model contains the right-handed neutrinos 
 in addition to the MSSM particles. 
Imposing the $\text{U(1)}_{L}^{}$ symmetry,
 the Majorana mass term of the right-handed neutrinos 
 in the superpotential is forbidden.
\begin{table}[h]
\caption{Additional matter in radiative induced neutrino
 mass model I.\label{table:model2}}
\begin{center}
\begin{tabular}{|c||c|c|c|c|}\hline
Superfields&$\text{SU(3)}_{\text{C}}^{}$&$\text{SU(2)}_{\text{L}}^{}$&$\text{U(1)}_{Y}^{}$&$\text{U(1)}_{L}^{}$\\\hline
$N_i^c$&$\pmb{1}$&$\pmb{1}$&$0$&$-1$\\ \hline
\end{tabular}
\end{center}
\end{table}
Then the superpotential is given by 
\begin{equation}
W_{I}=W_{\text{MSSM}}^{}+ y^\nu_{ij}N_i^cL_j\cdot H_2^{}\;.
\end{equation}
As mentioned above, the lepton number is kept as a good quantum number in the 
superpotential.
In our framework, the lepton number violation occurs only in the 
soft SUSY breaking terms induced by $F_{X_L}$.
Considering here the spurion possessing the lepton number $2$,
 which is denoted as $X_{L_2}$, 
 we can 
 write the soft SUSY breaking Lagrangian as
\begin{align}
\mathcal{L}_{\text{Model-I}}^{}=&\mathcal{L}_{\text{MSSM-}\not{L}}^{\text{soft}}
+\int d^4\theta \left\{
	\left(
	\frac{(c_N^{})^{ij}}{M^2}X^{\dagger}X
	+\frac{(\hat{c}_N^{})^{ij}}{M^2}X_{L_2}^{\dagger}X_{L_2}^{}
	\right)N_i^{c\dagger}N_j^c \right\}\nonumber\\
	&	+\int d^2\theta\left\{
		\frac{(a_{N}^{})^{ij}}{M}XN_i^cL_j\cdot H_2^{}
+(\hat{b}_N^{})^{ij} X_{L_2} N_i^cN_j^c
		+\text{h.c.}
		\right\}\;.
\end{align}
The $A$-term can be parametrized
 as $(a_{N}^{})_{ij}^{}=(\bar{a}_{N}^{})_{ij}^{}y^{\nu}_{ij}$. 
Then, 
 the right-handed sneutrino mass matrix is given by 
\begin{equation}
\tilde{m}^2_{\tilde{\nu}_R}\simeq 
\begin{pmatrix}
\frac{{c}_N}{{M}^2}|F_{X}|^2+\frac{\hat{c}_N}{{M}^2}|F_{X_{L_2}}|^2
 & \hat{b}_N F_{X_{L_2}}  \\
\hat{b}_N^* F_{X_{L_2}}^* &
 \frac{{c}_N}{{M}^2}|F_{X}|^2+\frac{\hat{c}_N}{{M}^2}|F_{X_{L_2}}|^2 
\end{pmatrix}\; 
\label{sNRmatrix}
\end{equation}
in the basis of $(\tilde{\nu}_R^*, \tilde{\nu}_R^{})$. 
Lepton number is violated by non-vanishing value of 
 $\hat{b}_NF_{X_{L_2}}$. 

\begin{figure}
\begin{center}
\begin{tabular}{cc}
\includegraphics[scale=0.6]{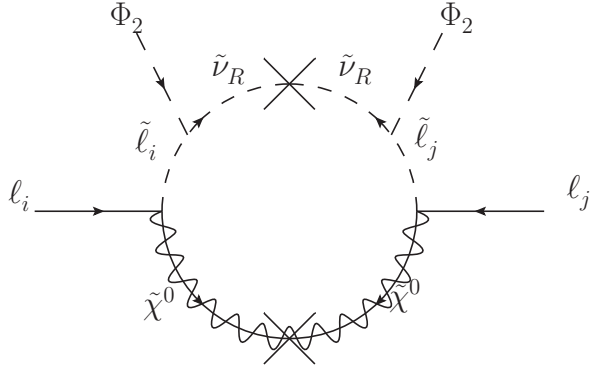}
\end{tabular}
\end{center}
\caption{Loop diagrams relevant to 
 the left-handed Majorana neutrino mass. 
}
\label{LoopMaj2}
\end{figure}

In this model, 
the Majorana mass terms for the neutrinos do not appear at the tree level.
It is naively expected that the light neutrinos are Dirac type.
However the right-handed sneutrino mass term breaks lepton number
and dimension five lepton number violating operator is induced 
through one-loop diagram shown in the Fig.~\ref{LoopMaj2}, 
as
\begin{equation}
\mathcal{L}_{\text{LNV}}^{}=
(C_L^{})_{ij}(\ell_i^{}\cdot \Phi_2^{})(\ell_j^{}\cdot \Phi_2^{})\;,
\label{37}
\end{equation}
where $(C_L)_{ij}$ is determined by sneutrino and SUSY breaking 
 mass parameters. 
The right-handed Majorana mass terms might be
 induced at the higher loop level,
 but they are suppressed much stronger.
Taking account of 
 the $n_N^{}$ generation ($n_N\leq 3$) of the right-handed neutrinos, 
 the $(3+n_N^{})\times (3+n_N^{})$ neutrino mass matrix is induced 
 after the electroweak symmetry breaking as
\begin{equation}
m_{\nu}^{}\simeq 
\begin{pmatrix}
m_{LL}^{}&m_D^{}\\
m_D^{}& 0 
\end{pmatrix}\;,
\label{310}
\end{equation}
where $m_{LL}^{}$ and 
 $m_{D}^{}$ are $3\times 3$ matrix and
 $n_N^{}\times 3$ matrix, respectively. 
$m_D$ is the Dirac mass matrix 
 as 
\begin{equation}
(m_D^{})^{ij}=(y_\nu^{})^{ij}\langle \Phi_2^{0}\rangle\;.
\end{equation}
On the other hand, 
 $m_{LL}$ is induced from dimension five operator in Eq.(\ref{37}), 
 which is given by  
\begin{equation}
(m_{LL}^{})^{ij}\sim
\frac{g^2}{(4\pi)^2}
\frac{\langle\Phi_2^0\rangle^2
m_{\tilde{\chi}^0_a}^{}\Gamma_{N}^{(a)}\Gamma_{N}^{(a)}
(A_{\nu}^{})_{ki}^{}(A_{\nu}^{})_{nj}^{}(\hat{b}_N)_{kn}
 F_{X_{L_2}}}
{\tilde{m}^{6}}\; ,
\label{eq:mLL}
\end{equation}
where 
 $(A_\nu)_{ij} = \frac{(\bar{a}_N^{})_{ij}y^{\nu}_{ij}}{{M}}F_{X}$, 
 and  
 $\tilde{m}$ is the diagonal element of 
 right-handed 
 sneutrino mass matrix in Eq.~(\ref{sNRmatrix}). 
Notice that 
 lepton number violating squared mass component, 
 $(\hat{b}_N^{})_{ij}F_{X_{L_2}}$,  
 plays a crucial role of generating Majorana 
 masses of light neutrinos, and  
 Majorana nature disappear as 
 $(\hat{b}_NF_{X_{L_2}}^{})\to 0$ (and also $\bar{a}_N\to 0$ as vanishing $A$-term). 
The coupling $\Gamma_{N}^{(a)}$ is defined as 
\begin{equation}
\Gamma_{N}^{(a)}=\sqrt{2}\left(\frac{1}{2}(U_N^*)_{a2}^{}-\frac{1}{2}\tan\theta_W(U_N^*)_{a1}^{}\right)\;,
\end{equation}
where the unitary matrix $U_N^{}$ is a neutralino mixing matrix as
\begin{equation}
\tilde{\chi}_a^{}=(U_N^{})_{ai}\tilde{\psi}_i^0\;,\quad \tilde{\psi}_i^{}=(\tilde{B}^0,\tilde{W}^0, \tilde{H}_1^0, \tilde{H}_2^0)\;.
\end{equation}
We then obtain radiative Majorana neutrino mass through 
 the effective dimension five operator.

Choosing the structure of the
 $y_\nu^{}$, $\bar{a}_N$, and $(\hat{b}_NF_{X_{L_2}})$,
 we can obtain 
 Dirac-, pseudo-Dirac-, or Schizophrenic-\cite{Allahverdi:2010us,Werner}
 types of light neutrinos.  
 For example, 
 the Dirac-type 
 neutrino is obtained when 
 $\bar{a}_N=0$ or $(\hat{b}_NF_{X_{L_2}})=0$. 
As for pseudo-Dirac-type neutrinos, 
 there are constraints for neutrino absolute mass,
 $\bar{a}_N$, and $(\hat{b}_NF_{X_{L_2}}^{})$, 
 since neutrino oscillation experiments suggest
 no large mixing to the sterile neutrinos.
The induced Majorana neutrino $m_{LL}^{}$ should be suppressed
 much smaller than the mass splitting of
 the active neutrinos, {\it i.e.},
 $m_D^{}m_{LL}^{}\ll\Delta m_{\odot}^2,\;\Delta m_{A}^2$, 
 which implies 
 \begin{equation}
 \frac{m_{\nu}^3m_{\tilde{\chi}^0}\bar{a}_N^2\hat{b}_N^{}F_X^2F_{X_{L_2}}}{M^2\tilde{m}^6}\ll \mathcal{O}(10^{-19}\text{GeV}^2)\;.
 \end{equation}
 For example,
 $m_{\tilde{\chi}^0}\simeq \tilde{m}=500\ \text{GeV}$,
 $m_{\nu}= 0.05\ \text{eV}$, and 
$(\hat{b}_N^{}F_{X_{L_2}})=100\ \text{GeV}^2$ are taken,
 the bound on the $A$-term is read
 as $({a}_N^{}F_X/M) \ll \mathcal{O}(10^{2}\text{GeV})$.

Next example is Schizophrenic-type
 neutrinos, which is realized with the 
 specific structure of 
 $y^{\nu}$, 
 $\bar{a}_N$, and $(\hat{b}_NF_{X_{L_2}}^{})$. 
For simplicity,
 let us consider one right-handed neutrino generation case ($n_N=1$).
With the following parameters, 
\begin{eqnarray}
&&y^{\nu}=
\begin{pmatrix}
2.7\times 10^{-14}&2.7\times 10^{-14}&2.7\times 10^{-14}
\end{pmatrix}\;,\nonumber \\
&&(a_N F_{X}/M)=\begin{pmatrix}
0&68&-68
\end{pmatrix}[\text{GeV}]\;,
\end{eqnarray}
$m_{\tilde{\chi}^0_a}=M=500\;\text{GeV}$, 
 and $(\hat{b}_N^{}F_{X_{L_2}})=100\ \text{GeV}^2$, 
the $4\times 4$ neutrino mass matrix becomes 
\begin{equation}
m_{\nu}=
\begin{pmatrix}
0&0&0&0.005\\
0&0.025&-0.025&0.005\\
0&-0.025&0.025&0.005\\
0.005&0.005&0.005&0
\end{pmatrix}[\text{eV}]
\end{equation}
which has the same structure as Eq.~(4) with $m_1^{}=0$ in 
 Ref.~\cite{Werner}. 
The mass matrix gives the tri-bimaximal neutrino mixing with massless
lightest neutrino.
With this parameter set, the second lightest neutrino is a Dirac type and the heaviest one 
is Majorana type.

Notice that 
 in the case of 
 extended gauge symmetry with $U(1)_{B-L}$
 as in Ref.\cite{Allahverdi:2010us},
 Majorana mass
  $m_{RR}^{}$
  is also induced through one-loop diagram.   
Where right-handed sneutrino and $B-L$ gaugino
 are propagating in internal lines, and 
 which can make a TeV-scale seesaw work.

\subsection{Loop induced Majorana neutrino mass model II}

The second model also contains 
 right-handed neutrinos, 
 however, 
 we assign 
 the $\text{U(1)}_{L}$-charge of $N^c$ 
 as quit different from conventional models as shown in 
 Table \ref{table:model3}.
We introduce three kinds of 
 spurion fields $X$, $X_{L_2}$, and $X_{L_6}$,
 which have lepton number $0$, $2$, and $6$, 
 respectively. 
\begin{table}[h]
\caption{Additional matter in radiative induced neutrino
 mass model II.\label{table:model3}}
\begin{center}
\begin{tabular}{|c||c|c|c|c|}\hline
Superfields&$\text{SU(3)}_{\text{C}}^{}$&$\text{SU(2)}_{\text{L}}^{}$&$\text{U(1)}_{Y}^{}$&$\text{U(1)}_{L}^{}$\\\hline
$N_i^c$&$\pmb{1}$&$\pmb{1}$&$0$&$-3$\\ \hline
\end{tabular}
\end{center}
\end{table}
Note that 
 Yukawa interaction of $N_i^cL_i^{}H_2^{}$ is 
 forbidden by $U(1)_{L}$ symmetry. 
The soft SUSY breaking terms
 replaced to 
 the right-handed neutrino 
 are generated from 
\begin{align}
\mathcal{L}_S^{}=&\mathcal{L}_{\text{MSSM-}\not{L}}^{\text{soft}}
+\int d^4\theta \left\{
	\left(
	\frac{(c_N^{})^{ij}}{M^2}X^{\dagger}X
	+\frac{(\tilde{c}_N^{})^{ij}}{M^2}X_{L_2}^{\dagger}X_{L_2}^{}
	+\frac{(\tilde{c}_N^{})^{ij}}{M^2}X_{L_6}^{\dagger}X_{L_6}^{}
	\right)N_i^{c\dagger}N_j^c
	\right\}\;
	\nonumber\\
	&+
	\int d^2\theta\left\{
	\frac{(\tilde{b}_N^{})^{ij}}{M}X_{L_6} M N_i^cN_j^c
	+\frac{(a_{N}^{})^{ij}}{M}X_{L_2} N_i^cL_i^{}\cdot H_2^{}
		+\text{h.c.}
		\right\}\;.
\end{align}
We assume that the right-handed supersymmetric 
 Majorana mass term of
 $M_{ij}\bar{\nu}_{Ri}\bar{\nu}^c_{Rj}$ is induced by a 
 mechanism associated with $U(1)_{B-L}$ breaking. 
Notice that 
 Dirac neutrino mass is forbidden, while 
 Majorana neutrino mass is allowed. 
Majorana masses of $m_{LL}$ are induced 
 through the one-loop diagram as Fig.1. 
After the electroweak symmetry breaking,
the $(3+n_N)\times (3+n_N)$ neutrino mass matrix is induced as
\begin{equation}
m_{\nu}^{}\simeq 
\begin{pmatrix}
m_{LL}^{}& 0\\
 0 & M 
\end{pmatrix}\;,
\end{equation}
where the formula for $m_{LL}^{}$ is given 
in Eq.~(\ref{eq:mLL}).
Let us 
 take a suitable parameter set for $n_N=2$ case, for example, 
\begin{equation}
(a_N F_{X}/M)=
\begin{pmatrix}
0&68&-68\\
-23&-23&-23
\end{pmatrix}[\text{GeV}]\;,
\end{equation}
with $m_{\tilde{\chi}_a^0}=M=500\;\text{GeV}$ 
 and $(\hat{b}_N^{}F_{X_{L_2}})=100\ \text{GeV}^2$, 
 we can obtain neutrino masses and mixing angles as 
 \begin{equation}
	m_1=0.0\text{eV}\;,\quad m_2=0.0088\text{eV}\;,\quad m_3=0.05\text{eV}\;,
 \end{equation}
 \begin{equation}
	\sin^22\theta_{A}=1.0\;,\quad \tan^2\theta_{\odot}=0.5\;, \quad \sin^2\theta_{13}=0.0\;,
 \end{equation}
which can explain the neutrino oscillation data with 
 tri-bimaximal mixing.

We should comment that, 
 if there exist $U(1)_{B-L}$ gauge symmetry,
 Dirac mass is also induced through one-loop diagram. 
Where, by $\tilde{b_N}$ coupling, sneutrino mass is modified,
 but its effect is negligible comparing to SUSY Majorana mass.

\subsection{Application to SUSY Zee-Babu model}

We can also apply our setup to known models, where the lepton number
violation is required for generating neutrino masses.
A famous example is Zee-Babu model\cite{zee-babu}, and trial of SUSY extension of
the model has been first suggested in Ref.~\cite{Aoki:2010ib}.
This model  has extra 
 two pairs of the $SU(2)$-singlet fields
 (and their  hypercharges are $\pm 1$ and $\pm 2$, respectively)
 in addition to the MSSM as Table 4. 
\begin{table}[h]
\caption{Additional matter in SUSY Zee-Babu Model\cite{Aoki:2010ib}. 
\label{table:model1}}
\begin{center}
\begin{tabular}{|c||c|c|c|c|}\hline
Superfields&$\text{SU(3)}_{C}^{}$&$\text{SU(2)}_{L}^{}$&$\text{U(1)}_{Y}^{}$&$\text{U(1)}_{\text{L}}^{}$\\\hline
$\Omega_{\pm}^{}$&$\pmb{1}$&$\pmb{2}$&$\pm 1$&$\mp 2$\\
$K_{\pm}^{}$&$\pmb{1}$&$\pmb{2}$&$\pm 2$&$\mp 2$\\ \hline
\end{tabular}
\end{center}
\end{table}
In this model, the superpotential is written as 
\begin{equation}
W=W_{\text{MSSM}}^{}+f_{ij}L_i^{}\cdot L_j^{}\Omega_+^{}+g_{ij}^{}E_i^cE_j^cK_{-}
+\mu_{\Omega}^{}\Omega_+^{}\Omega_-^{}
+\mu_{K}^{}K_+^{}K_-^{}\;.
\end{equation}
Let us introduce spurion fields, $X$ and $X_{L_2}$, 
 which have lepton number $0$ and $2$, respectively. 
Then, interactions between these spurion fields and the scalar
 components of the visible sector fields are given by 
\begin{align}
\mathcal{L}_S^{}=&\mathcal{L}_{\text{MSSM}}^{} 
+\int d^4\theta
\left\{
	\left(\frac{c_{\Psi}^{}}{M^2}X^{\dagger}X
	+\frac{\tilde{c}_{\Psi}^{}}{M^2}X_{L_2}^{\dagger}X_{L_2}^{}
	\right)\Psi^{\dagger}\Psi
	\right\}
	\nonumber\\
	&+\int d^2\theta
	\left\{
	b_{\Omega}^{}X\Omega_+\Omega_{-}
	+b_{K}^{}XK_+K_{-}
		+\frac{(a_{LL}^{})^{ij}}{M}XL_i\cdot L_j\Omega_+
		+
		\frac{(a_{EE}^{})^{ij}}{M}XE^c_iE^c_jK_-
		\right.\nonumber\\
		&\left.\phantom{Spa}+
		\frac{\tilde{a}_{-}^{}}{M}X_{L_2}
K_-\Omega_+\Omega_+
		+\text{h.c.}
		\right\}\;,
\end{align}
where we $\Psi$ represents $\Psi =\Omega_+, \Omega_-, K_+, K_-$. 
Notice that 
 the lepton number is softly broken by two units
 when the SUSY is broken by the
 $F$-term of the spurion, $X_{L_2}$. 

\begin{figure}
\begin{center}
\includegraphics[scale=0.6]{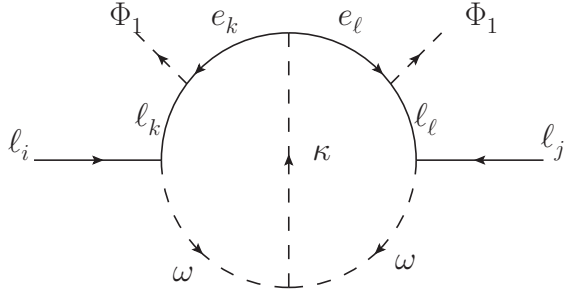}
\end{center}
\caption{The diagram for two-loop neutrino mass generation.}
\label{TwoLoopMaj}
\end{figure} 
The neutrino mass in the model is induced through
 two-loop diagram in Fig. 2 which is the same as 
 non-SUSY Zee-Babu model\cite{zee-babu}. 
The neutrino mass in the model is evaluated as\cite{zee-babu,Aoki:2010ib},
\begin{equation}
(m_{\nu})_{ij}^{}=
\left(\frac{1}{16\pi^2}\right)^2
\frac{16\mu_B^{}f_{ik}m_{\ell_k}^{}g_{kl}m_{\ell_l}^{}f_{jl}^{}}{m_{\kappa}^2}I\left(\frac{m_{\kappa}^2}{m_{\omega}^2}\right)\;,
\end{equation}
where $m_{\kappa}^{}$ and $m_{\omega}^{}$ are the masses of the scalar components of $K_-$ and $\Omega_+$, 
$I(x)$ is the loop function given by\cite{loopfuncZB}
\begin{equation}
I(r)=-\int_0^1dx\int_0^{1-x}dy\frac{1}{x+(r-1)x+y^2}\ln\frac{y(1-y)}{x+ry}\;,
\end{equation}
and the lepton number violating coupling $\mu_B^{}$ is 
\begin{equation}
\mu_B^{}=\frac{\tilde{a}_-^{}\langle F_{X_{L_2}^{}}^{}\rangle}{M}
\;.
\end{equation}
In this setup, the lepton number is a good quantum number 
with softly breaking parameter.
Therefore,  our suggesting setup can cure the model 
 and we can obtain accurately modified SUSY Zee-Babu model. 
Since the lepton number symmetry is broken {\it softly}, 
 lepton number violating processes
 can be strongly suppressed comparing to 
 lepton number hardly breaking model\cite{Aoki:2010ib}.
 For example, a lepton number violating 
 cross section of $\mu\to e \nu \nu$ is tiny 
 since it is induced through two-loop diagram in our 
 {\it softly} breaking setup, 
 while this process appears in one-loop diagram 
 in {\it hardly} breaking setup. 

\section{Conclusion}

We have suggested new setup where 
 SUSY breaking spurion $F$-term possesses lepton number.
This setup not only modifies sparticle mass spectra 
 but also realizes several new models, 
 where neutrino mass 
 is
 naturally
 induced by radiative corrections. 
We have suggested two new models; the first one is 
 (i): pseudo-Dirac/Schizophrenic neutrino model, and 
 the second one is  
 (ii): pure Majorana neutrino model.
We have also shown  
 this setup can naturally apply to the supersymmetric 
 Zee-Babu model.
The necessary lepton number violating couplings 
 are induced in the soft SUSY breaking sector, 
 while the lepton number is kept in the superpotential.


\vspace{1cm}

{\large \bf Acknowledgments}\\

\noindent
We thank S.~Yamaguchi and T.~Yamashita 
 for helpful discussions. 
This work is partially supported by the Ministry of Education and
Science, Scientific Grant  No. 20105002.
We would like to thank the high energy theory group of Osaka
University, where discussions during the
"International workshop" of the high energy theory group of Osaka
university at Ohmi-Hachiman, 3-5 Dec. 2010
opened up this work.


\end{document}